\documentclass[prb,10pt,aps,twocolumn,showpacs,showkeys,citeautoscript]{revtex4-1}

\usepackage{graphicx}
\usepackage{amsmath}
\usepackage{amssymb}
\usepackage{bm}

\newcommand{\ud}{\,{\mathrm d}}
\newcommand{\uiiint}{\int\!\!\!\!\int\!\!\!\!\int}
\newcommand{\ueex}{e_\mathrm{EX}}
\newcommand{\uea}{e_\mathrm{A}}
\newcommand{\uems}{e_\mathrm{MS}}
\newcommand{\uez}{e_\mathrm{Z}}
\newcommand{\uMS}{\widetilde{M}_\mathrm{S}}
\newcommand{\uMx}{\widetilde{M}_\mathrm{X}}

\newcommand{\uMy}{\widetilde{M}_\mathrm{Y}}

\newcommand{\uMz}{\widetilde{M}_\mathrm{Z}}

\newcommand{\uImr}{I_\mathrm{m}}
\newcommand{\uIm}{\widetilde{I}_\mathrm{m}}

\newcommand{\uI}{\widetilde{I}}

\newcommand{\uIer}{I_\mathrm{e}}
\newcommand{\uIe}{\widetilde{I}_\mathrm{e}}
\newcommand{\uIkr}{I_\mathrm{k}}
\newcommand{\uIk}{\widetilde{I}_\mathrm{k}}
\newcommand{\udx}{\widetilde{d}_\mathrm{X}}
\newcommand{\udy}{\widetilde{d}_\mathrm{Y}}
\newcommand{\udz}{\widetilde{d}_\mathrm{Z}}
\newcommand{\uAx}{\widetilde{A}_\mathrm{X}}
\newcommand{\uAy}{\widetilde{A}_\mathrm{Y}}

\newcommand{\umIx}{\widetilde{m}_\mathrm{X}^{(1)}\!}

\newcommand{\umIy}{\widetilde{m}_\mathrm{Y}^{(1)}\!}

\newcommand{\umII}{\widetilde{\vec{m}}^{(2)}\!}
\newcommand{\umIIx}{\widetilde{m}_\mathrm{X}^{(2)}\!}

\newcommand{\umIIy}{\widetilde{m}_\mathrm{Y}^{(2)}\!}

\newcommand{\umIIz}{\widetilde{m}_\mathrm{Z}^{(2)}\!}

\newcommand{\uxq}{x_\vec{q}}
\newcommand{\uyq}{y_\vec{q}}
\newcommand{\uzq}{z_\vec{q}}
\newcommand{\uuq}{u_\vec{q}}
\newcommand{\uxqp}{x_{\vec{q}^\prime}}
\newcommand{\uyqp}{y_{\vec{q}^\prime}}
\newcommand{\uzqp}{z_{\vec{q}^\prime}}
\newcommand{\uuqp}{u_{\vec{q}^\prime}}
\newcommand{\uxqmqp}{x_{\vec{q}-\vec{q}^\prime}}
\newcommand{\uyqmqp}{y_{\vec{q}-\vec{q}^\prime}}
\newcommand{\uzqmqp}{z_{\vec{q}-\vec{q}^\prime}}
\newcommand{\uuqmqp}{u_{\vec{q}-\vec{q}^\prime}}

\newcommand{\uLe}{L_\mathrm{EX}}
\newcommand{\uLeO}{L_{0}}

\newcommand{\udet}{\mathrm{det}\,}

\let\oldhat\hat
\renewcommand{\vec}[1]{{\bm{#1}}}
\renewcommand{\hat}[1]{\oldhat{{\bm{#1}}}}
\newcommand{\vecT}[1]{{\bm{#1}}^{\!\intercal}\!}
\newcommand{\mat}[1]{{\bm{#1}}}
\newcommand{\matT}[1]{{\bm{#1}}^{\!\intercal}\!}
\newcommand{\matI}[1]{{\bm{#1}}^{\!-\!1}\!}

\begin{document}

\title{Third-order effect in magnetic small-angle neutron scattering by a spatially inhomogeneous medium}

\author{Konstantin L. Metlov}
\affiliation{Donetsk Institute for Physics and Technology NAS, Donetsk, Ukraine 83114}
\email{metlov@fti.dn.ua}
\author{Andreas Michels}
\affiliation{Physics and Material Science Research Unit, University of Luxembourg, 162A Avenue de la Fa\"iencerie, L-1511 Luxembourg, Grand Duchy of Luxembourg}

\date{\today}

\begin{abstract}
Magnetic small-angle neutron scattering (SANS) is a powerful tool for investigating nonuniform magnetization structures inside magnetic materials. Here, considering a ferromagnetic medium with weakly inhomogeneous uniaxial magnetic anisotropy, saturation magnetization, and exchange stiffness, we derive the second-order (in the amplitude of the inhomogeneities) micromagnetic solutions for the equilibrium magnetization textures and compute the corresponding magnetic SANS cross sections up to the next, third order. We find that in the case of perpendicular scattering (the incident neutron beam is perpendicular to the applied magnetic field) if twice the cross section along the direction orthogonal to both the field and the neutron beam is subtracted from the cross section along the field direction, the result has only a third-order contribution (the lower-order terms are canceled). This difference does not depend on the amplitude of the exchange inhomogeneities and provides a separate gateway for a deeper 
analysis of the sample's magnetic structure. We derive and analyze analytical expressions for the dependence of this combination on the scattering-vector magnitude for the case of spherical Gaussian inhomogeneities.
\end{abstract}



\maketitle

\section{Introduction}

Magnetic small-angle neutron scattering (SANS) is an important tool for the analysis of magnetic structures on the nanoscale\cite{M14_review}. Traditional scalar magnetometry, for example, only measures the sample's total magnetic moment and has no spatial resolution. Magnetic force microscopy is sensitive to the spatial features of the magnetization only in the near-vicinity of the sample's surface and is also prone to disturbing the magnetic structure during the measurement. Optical magnetometry is either also only surface-sensitive (such as in Kerr microscopy) or is applicable only to optically transparent magnets (such as in Faraday microscopy). Magnetic SANS complements these techniques by permitting analysis of the sample's magnetic structure throughout the volume, even in nontransparent materials, while also being sensitive to the spatial arrangement of the magnetization.

The analysis of magnetic SANS cross sections is closely interwoven\cite{KSF63} with the continuum theory of micromagnetics\cite{brown1963micromagnetics}. This is because, unlike nonmagnetic nuclear SANS (which is sensitive to nanoscale density and compositional fluctuations), magnetic SANS cross section images are formed by the distribution of the magnetic moments within the sample. These magnetic moments are influenced by magnetic material inhomogeneities, but, due to their mutual interaction, do not follow the inhomogeneities exactly. Thus, in order to understand magnetic SANS cross sections, one must also understand the process of magnetic-structure formation and its dependence on the external magnetic field, which is the subject of micromagnetic theory.

Currently, the interpretation of magnetic SANS cross sections of heterogeneous multiphase magnets with small inhomogeneities of the saturation magnetization and the magnetic anisotropy is based on a second-order (in the inhomogeneities amplitude) theory, \cite{michels2013} which has its origin in the theory of the approach to magnetic saturation\cite{kronm2003micromagnetism}. The latter stems from the works of Schl\"omann \cite{S71}, which, in turn, is a follow up on the work by N\'eel \cite{N45}.

The motivation for the present study is to probe the limits of the second-order magnetic SANS theory \cite{michels2013} by looking for prominent third-order effects. Specifically, those, which are not masked by the second-order ones. One such effect---a central result of this work---is pinpointed at the end of the paper (Sec.~\ref{sec:3rdorder}). An attempt was made to include all the interactions which are common in micromagnetics. In particular, our solution for the micromagnetic problem of weakly inhomogeneous magnets includes inhomogeneous exchange interaction, which is irrelevant for the problem of the approach to magnetic saturation and for the second-order SANS theory. Also, the present theory explicitly includes a weak, fluctuating random-axis uniaxial anisotropy and full three-dimensional anisotropy-direction averaging. 

The paper is organized as follows. Sec.~\ref{sec:sanscross} introduces the magnetic SANS cross sections, Sec.~\ref{sec:micro} details the solution of the micromagnetic problem in Fourier space, Sec.~\ref{sec:averag} provides a discussion of the defect distribution and averaging procedures, the second and higher-order SANS cross sections are, respectively, discussed in Secs.~\ref{sec:2ndorder} and \ref{sec:3rdorder}, and Sec.~\ref{sec:sum} summarizes the main results of this study.

\section{Magnetic SANS cross sections}
\label{sec:sanscross}

The current theoretical and experimental understanding of magnetic SANS of bulk ferromagnets has recently been summarized \cite{M14_review}. The quantity of interest---the differential SANS cross section $\frac{\ud \Sigma}{\ud \Omega}$---is related to the Fourier transform of the Cartesian components of the magnetization vector field $\widetilde{\vec{M}}=\{\widetilde{M}_\mathrm{X},\widetilde{M}_\mathrm{Y},\widetilde{M}_\mathrm{Z}\}$. In particular, the total unpolarized nuclear and magnetic SANS cross section is \cite{MW08}
\begin{eqnarray}
\frac{\ud \Sigma^\perp}{\ud \Omega} & = &
8\pi^3 V b_\mathrm{H}^2\left[ \frac{|\widetilde{N}|^2}{b_\mathrm{H}^2} +
|\uMx|^2 + |\uMy|^2 \cos^2{\alpha} +\right. \nonumber \\
& & \left. |\uMz|^2 \sin^2 \alpha - 2 \operatorname{Re}(\overline{\uMy} \uMz)\sin\alpha\cos\alpha \right]
\label{eq:unpolarizedperp} , \\
\frac{\ud \Sigma^\parallel}{\ud \Omega} & = &
8\pi^3 V b_\mathrm{H}^2\left[ \frac{|\widetilde{N}|^2}{b_\mathrm{H}^2} +
|\uMx|^2 \sin^2{\beta} + |\uMy|^2 \cos^2{\beta} +\right. \nonumber \\
& & \left. |\uMz|^2 - 2 \operatorname{Re}(\overline{\uMy} \uMx)\sin\beta\cos\beta \right] 
\label{eq:unpolarizedpar},
\end{eqnarray}
where the first expression refers to the perpendicular scattering geometry, for which $\vec{q}^\perp=q \{0, \sin\alpha, \cos\alpha \}$, and the second equation relates to the parallel geometry, with $\vec{q}^\parallel=q \{\cos\beta, \sin\beta, 0 \}$; $V$ is the scattering volume, $b_\mathrm{H} = 2.9 \times 10^8 \mathrm{A}^{-1}\mathrm{m}^{-1}$, $\widetilde{N}(\vec{q})$ is the nuclear scattering amplitude, $\operatorname{Re}$ stands for taking the real part of a complex number and over-bar for its complex conjugate. The cross section is generally measured in units of $\mathrm{cm}^{-1}\mathrm{sr}^{-1}$.

Fourier transforms (distinguished by tildes above the symbol) are defined for a representative cube of the material with dimensions $L \times L \times L$. Most of the time we shall work with discrete transforms (the continuous ones correspond to the limit $L\rightarrow\infty$):
\begin{eqnarray}
 \widetilde{X}(\vec{q}) & = & \frac{1}{V} \uiiint_{V} X(\vec{r})\, e^{-\imath \vec{q}\vec{r}}  \ud^3 \vec{r}, \\
 X(\vec{r}) & = & \sum\nolimits_{\vec{q}} \widetilde{X}(\vec{q})\, e^{\imath \vec{q}\vec{r}},
\end{eqnarray}
where $\imath=\sqrt{-1}$, $V = L^3$ is the representative cube volume and $X$ is a quantity which is defined inside the volume. The components of $\vec{q}=\{q_\mathrm{X}, q_\mathrm{Y}, q_\mathrm{Z}\}$ take on all values which are integer multiples of $2\pi/L$. Note that unlike Ref.~\onlinecite{MW08} our definition of the forward Fourier transform carries a prefactor of $1/V$, so that the Fourier transform has the same dimension as the transformed quantity. This, however, renders the prefactor in the expressions for the cross sections, Eqs.~(\ref{eq:unpolarizedperp}) and (\ref{eq:unpolarizedpar}), also slightly different.

In order to get rid of the nuclear scattering $|\widetilde{N}|^2$, the cross sections are typically split into the residual and the magnetic parts, $\Sigma = \Sigma_\mathrm{res} + \Sigma_\mathrm{M}$, where the residual part corresponds to the magnet in the fully saturated state. Assuming that a large saturating external magnetic field is directed along the $Z$-axis,
\begin{eqnarray}
\frac{\ud \Sigma^\perp_\mathrm{res}}{\ud \Omega} & = &
8\pi^3 V b_\mathrm{H}^2\left[ \frac{|\widetilde{N}|^2}{b_\mathrm{H}^2} + |\uMS|^2 \sin^2 \alpha  \right] , \\
\frac{\ud \Sigma^\parallel_\mathrm{res}}{\ud \Omega} & = &
8\pi^3 V b_\mathrm{H}^2\left[ \frac{|\widetilde{N}|^2}{b_\mathrm{H}^2} + |\uMS|^2 \right],
\end{eqnarray}
where $\widetilde{M}_{\mathrm{S}}$ is the Fourier transform of the inhomogeneous saturation magnetization of the magnet, which is defined in the next section. The residual part can then be measured independently and subtracted from the measured total cross section at a lower field to yield the magnetic part:
\begin{eqnarray}
\frac{\ud \Sigma^\perp_\mathrm{M}}{\ud \Omega} & = &
8\pi^3 V b_\mathrm{H}^2\left[
|\uMx|^2 + |\uMy|^2 \cos^2{\alpha} + \right. \nonumber \\
& &  (|\uMz|^2-|\uMS|^2) \sin^2 \alpha - \nonumber \\
& & \left. 2 \operatorname{Re}(\overline{\uMy} \uMz)\sin\alpha\cos\alpha \right] \label{eq:sigmamagperp} , \\
\frac{\ud \Sigma^\parallel_\mathrm{M}}{\ud \Omega} & = &
8\pi^3 V b_\mathrm{H}^2\left[
|\uMx|^2 \sin^2{\beta} + |\uMy|^2 \cos^2{\beta}  + \right. \nonumber \\
& & \left. (|\uMz|^2-|\uMS|^2) - 2 \operatorname{Re}(\overline{\uMy} 
     \uMx)\sin\beta\cos\beta \right] . \nonumber \\ \label{eq:sigmamagpar}
\end{eqnarray}

Thus, in order to compute the magnetic SANS cross section one needs to know the Fourier components of the magnetization vector field inside the material. Their derivation is the main subject of the two following sections.

\section{Magnetization distribution in a weakly inhomogeneous magnet}
\label{sec:micro}

Consider an infinite magnet, whose saturation magnetization depends explicitly on the position vector $\vec{r}$,
\begin{equation}
 \label{eq:msi}
 M_S(\vec{r}) = M_0\left[1+\uImr(\vec{r})\right] ,
\end{equation}
where the magnitude of $\uImr(\vec{r})$ is a small quantity. We also assume that the spatial average $\langle \uImr(\vec{r}) \rangle=0$, so that $M_0= \langle M_S(\vec{r}) \rangle$ is the average saturation magnetization of the magnet.

If the representative volume contains a magnetic material, the equilibrium distribution of the magnetization vector $\vec{M}(\vec{r})$ is the solution of Brown's equations \cite{Brown_micromagnetics} at each point $\vec{r}$ in the volume,
\begin{equation}
[\vec{H}^\mathrm{eff},\vec{M}] = 0, \label{eq:Brown}
\end{equation}
where the square brackets denote the vectorial cross product. The effective field $\vec{H}^\mathrm{eff}(\vec{r})$ is defined as the functional derivative of the ferromagnet's energy-density functional $e$ over the magnetization vector field, 
\begin{equation}
\vec{H}^\mathrm{eff}(\vec{r}) = -\frac{1}{\mu_0} \frac{\delta e}{\delta \vec{M}} =
- \frac{1}{\mu_0} \left( \frac{\partial e}{\partial \vec{M}} - \frac{\partial}{\partial \vec{r}} \frac{\partial e}{\frac{\partial \vec{M}}{\partial \vec{r}}} \right).
\end{equation}

From the magnetic-units standpoint, following Aharoni, \cite{AharoniBook} we use the defining 
relation for the magnetic induction
\begin{equation}
 \vec{B}=\mu_0\,(\vec{H} + \gamma_\mathrm{B}\, \vec{M}), \label{eq:induction}
\end{equation}
which can be made valid in all systems of magnetic units by appropriately choosing the constants $\mu_0$ and $\gamma_\mathrm{B}$. For example, in the SI system $\mu_0$ is the permeability of vacuum and $\gamma_\mathrm{B}=1$, in the CGS system $\mu_0=1$ and $\gamma_\mathrm{B}=4\pi$.

The energy density $e$ represents our knowledge of the interactions in the magnetic material. Here, we include the effects of exchange, random uniaxial anisotropy, magnetostatic interaction, and
the influence of the uniform external magnetic field, so that the total energy density of the magnet can be written as a sum,
\begin{equation}
 e =  \ueex + \uea + \uems + \uez \label{Efull} .
 \end{equation}
Different interactions enter both the energy density and the effective field additively, so that
\begin{equation}
 \vec{H}^\mathrm{eff} =  \vec{H}_\mathrm{EX} + \vec{H}_\mathrm{A} + 
 \vec{H}_\mathrm{MS} + \vec{H}_\mathrm{Z} \label{Hfull},
\end{equation}
where $\vec{H}_Z$ is simply the external field.

The exchange interaction is deemed inessential in the theory of the approach to magnetic saturation\cite{S71,N45}, but SANS is sensitive to small spatial variations of the magnetization vector field, despite their negligible contribution to the total magnetization of the sample. That is why we have included the exchange interaction into consideration. Its energy density in a material with varying saturation magnetization is conventionally defined as
\begin{equation}
 \ueex  =  \frac{C(\vec{r})}{2}
       \sum_{i=X,Y,Z} \left(
       \vec{\nabla} \left[ \frac{M_i(\vec{r})}{M_S(\vec{r})} \right]
       \right)^2 ,
\end{equation}
where $C(\vec{r})$ is the exchange stiffness; $X$, $Y$, $Z$ are the labels of the Cartesian coordinate-system axes; $\vec{\nabla}=\{\partial/{\partial X}, \partial/{\partial Y}, \partial/{\partial Z}\}$ is the gradient operator. Using vector-calculus identities, it can be transformed into 
\begin{equation}
 e_\mathrm{EX} = \frac{\mu_0 \gamma_B \uLe^2(\vec{r})}{2}\left(
     -\left[\vec{\nabla} M_S(\vec{r})\right]^2 + 
     \!\!\!\!\!\sum_{i=X,Y,Z}\!\!\!\!\left[\vec{\nabla} M_i(\vec{r})\right]^2
 \right) , \label{eq:eEX}
\end{equation}
with the exchange length
\begin{equation}
 \uLe(\vec{r})=\sqrt{C(\vec{r})/[\mu_0 \gamma_B M_S(\vec{r})]}.
\end{equation}
The first term in Eq.~(\ref{eq:eEX}) vanishes under the variation of $\vec{M}$ and gives no contribution to the effective field. This is a manifestation of the fact that the exchange energy depends only on relative angles of the magnetization vectors and not on their magnitude. Thus,
\begin{equation}
 \vec{H}_\mathrm{EX} = \gamma_B \uLe^2(\vec{r}) \Delta \vec{M}(\vec{r}) +\gamma_B \vec{\nabla}\uLe^2(\vec{r})
 \vec{\nabla} \vec{M}(\vec{r}),
\end{equation}
where $\Delta$ is the Laplace operator and $\vec{\nabla} \vec{M}(\vec{r})$ is a matrix, whose rows are the gradients of the components of the vector $\vec{M}(\vec{r})$. Similarly to Eq.~(\ref{eq:msi}) we will now assume that the squared exchange length is weakly inhomogeneous 
\begin{equation}
\uLe^2(\vec{r}) = \uLeO^2 \left[1 + \uIer (\vec{r}) \right],
\end{equation}
where $\uIer$ is a small position-dependent quantity of the same order as $\uImr$ and $\uLeO$ is an average position-independent exchange length. In real materials, since the values of both $C$ and $M_\mathrm{S}$ are determined by the same quantum exchange interaction (and both grow as exchange becomes stronger), the value of $\uLeO$ displays little variation across a wide range of magnetic materials and is of the order of $5-10$~nm for most of them. Nevertheless, we shall keep track of the weak spatial dependence of $\uLe$ in this calculation.

The presence of uniaxial anisotropy creates the following energy density
\begin{equation}
 \uea = - k_\mathrm{U}(\vec{r})  \frac{[\vec{d}(\vec{r})\cdot\vec{M}(\vec{r})]^2}{M^2_\mathrm{S}(\vec{r})},
\end{equation}
where $k_\mathrm{U}(\vec{r})$ is the spatially inhomogeneous anisotropy constant, and $\vec{d}(\vec{r})$ is a unit vector along the local direction of the anisotropy axis. The corresponding effective field is
\begin{equation}
  \vec{H}_\mathrm{A}=\gamma_B Q(\vec{r}) [\vec{d}(\vec{r})\cdot\vec{M}(\vec{r})] \vec{d}(\vec{r}),
\end{equation}
where the dimensionless quality factor
\begin{equation}
 Q(\vec{r}) = 2 k_\mathrm{U}(\vec{r})/[\mu_0 \gamma_B M^2_S(\vec{r})] = \uIkr(\vec{r})
\end{equation}
is assumed to be small and of the same order as $\uImr$.

The magnetostatic energy density is
\begin{equation}
 \uems = -\frac{1}{2} \mu_0 (\vec{H}_\mathrm{D} \cdot \vec{M}),
\end{equation}
where $\vec{H}_\mathrm{D}$ is the magnetostatic (or demagnetizing) field. The expression for the latter in the static case with no macroscopic currents is simplest in Fourier representation. It follows \cite{herring51} from the expression of the magnetic induction (\ref{eq:induction}) with the internal field $\vec{H} = \vec{H}_Z + \widetilde{\vec{H}}_\mathrm{D}$ and Maxwell's equations $\nabla \times \vec{H}_\mathrm{D} = 0$ and $\nabla \cdot \vec{B} = 0$ that 
\begin{equation}
 \widetilde{\vec{H}}_\mathrm{D}=-\gamma_B \frac{\vec{q}(\vec{q}\cdot\widetilde{\vec{M}})}{q^2}\,\,\,
 \mathrm{for}\,\,\, q\neq0 . \label{eq:hdq}
\end{equation}
The {\em average} demagnetizing field $\widetilde{\vec{H}}_\mathrm{D}(0)$ is anti-parallel to the external field $\vec{H}_\mathrm{Z}$. Thus, we can add these fields as scalars $H = H_Z - |\widetilde{\vec{H}}_\mathrm{D}(0)|$. All the following results will be computed as functions of this internal field $H$ (which, in particular, contains the information about the shape of the sample) and not directly of the external field $H_Z$.

Having a simple expression for the demagnetizing field, Eq.~(\ref{eq:hdq}), suggests trying to solve Brown's equations, Eqs.~(\ref{eq:Brown}), directly in Fourier space. There is, however, a complication, since products of functions in real space become their convolutions in Fourier space. Thus, to simplify the expressions, let us introduce a shorthand notation for convolutions:
\begin{equation}
 X \otimes Y (\vec{q}) = \sum\limits_{\vec{q}^\prime} X(\vec{q}^\prime) Y(\vec{q}-\vec{q}^\prime),
\end{equation}
where the argument $\vec{q}$ on the left hand side (which sometimes will be omitted in the following text) is the argument of the whole convolution (not just of $Y$) and summation is carried out over all the values of $\vec{q}^\prime$. The algebra of convolutions is commutative ($X \otimes Y = Y \otimes X$), distributive ($X \otimes (Y + Z) = X \otimes Y + X \otimes Z$), and associative with respect to multiplication of a constant, $a (X \otimes Y) = (a X) \otimes Y = X \otimes (a Y)$, where $a$ is a constant. It also has an identity element (a product of Kronecker deltas in the discrete case or Dirac's delta functions in the continuous case), which we will denote as $\delta(\vec{q})$, so that $\delta \otimes X = X$. We will also sometimes specify functions in-line by underlining them, so that $\underline{q Z} \otimes Y$ is convolution of the function $X(\vec{q}) = q Z(\vec{q})$ with the function $Y(\vec{q})$.

Using this notation, we can now express Fourier representations of the effective-field terms:
\begin{eqnarray}
 \widetilde{\vec{H}}_\mathrm{EX} & = &
      - \gamma_B \widetilde{\uLe^2} \otimes \underline{q^2 \widetilde{\vec{M}}} \nonumber \\
   & &\qquad  
   - \gamma_B \underline{\vec{q} \widetilde{\uLe^2}} \otimes \underline{\vec{q}\times\widetilde{\vec{M}}} , \\
 \widetilde{\vec{H}}_\mathrm{A} & = & \gamma_B \widetilde{Q} \otimes \left(
 \widetilde{d}_\mathrm{X}\otimes\widetilde{M}_\mathrm{X}+
 \widetilde{d}_\mathrm{Y}\otimes\widetilde{M}_\mathrm{Y}+ \right.  \nonumber \\
 & & \qquad \left. \widetilde{d}_\mathrm{Z}\otimes\widetilde{M}_\mathrm{Z}\right)
 \otimes \widetilde{\vec{d}} , \\
 \widetilde{\vec{H}}_\mathrm{MS}+\widetilde{\vec{H}}_\mathrm{Z} & = & \vec{H}\delta 
      -\gamma_B \frac{\vec{q}[\vec{q}\cdot(\widetilde{\vec{M}}-\delta\widetilde{\vec{M}})]}{q^2},
\end{eqnarray}
where the cross ($\times$) denotes a direct product of two vectors (forming a matrix, having the products of the
left vector by each element of the right vector in the rows) and convolution of a vector with a matrix is like their normal product but with convolutions instead of multiplications.
The final subtraction, together with the condition that $\widetilde{\vec{M}}-\delta\widetilde{\vec{M}}\rightarrow 0$ as $\vec{q}\rightarrow 0$ is a mathematical trick, allowing not to pay further attention to the fact that the expression for the demagnetizing field, Eq.~(\ref{eq:hdq}), is valid only for $\vec{q}\neq 0$. This limiting condition is fulfilled, if $\langle \uImr(\vec{r}) \rangle= \uIm(0) =  0$, which is assumed from the start of this computation.

In completely homogeneous infinite isotropic magnets, the magnetization will always be uniform,  saturated, and aligned parallel to the external (and the internal) field (however small it is). In our weakly-inhomogeneous and weakly anisotropic case, there will be a small deviation from uniformity. Let us choose the coordinate system in such a way that the direction of the external field coincides with $Z$-axis, so that $\vec{H} = \{0 , 0, H\}$ and, using the magnitude of $\uImr$ as a small parameter, represent this weakly inhomogeneous magnetization via Taylor-series expansion
\begin{equation}
 \widetilde{\vec{M}} = \{0,0,M_0\}\delta + \widetilde{\vec{M}}^{(1)} + \widetilde{\vec{M}}^{(2)}+\ldots,
\end{equation}
where $\vec{M}^{(i)}$ contains the terms of the order $i$ in $\uImr$.

Due to the constraint $\vec{M}^2(\vec{r})=M_\mathrm{S}^2(\vec{r})$ there are only two independent components of $\vec{M}$. Considering $M_\mathrm{X} = M_\mathrm{X}^{(1)}+M_\mathrm{X}^{(2)}$ and $M_\mathrm{Y} = M_\mathrm{Y}^{(1)}+M_\mathrm{Y}^{(2)}$ independent and small, the expansion of the constraint up to the second order allows us to express the remaining component of $\vec{M}$ in real space as
\begin{equation}
M_\mathrm{Z}=M_0+M_0\uImr - \frac{(M_\mathrm{X}^{(1)})^2 + (M_\mathrm{Y}^{(1)})^2}{2M_0} . \label{eq:mzmag}
\end{equation}
Rendering products as convolutions in Fourier space and introducing the dimensionless magnetization vector $\vec{m}=\vec{M}/M_0$ we get
\begin{equation}
    \widetilde{m}_\mathrm{Z}=\delta + \uIm - F_\mathrm{Z}, \label{eq:mzdir}
\end{equation}
where
\begin{equation}
 F_\mathrm{Z}= \frac{\umIx \otimes \umIx + \umIy \otimes \umIy}{2}. \label{eq:Fz}
\end{equation}

Brown's equations (of which only two are independent) in Fourier space also contain convolutions. For example, the first one of them reads
\begin{equation}
 \widetilde{H}^\mathrm{eff}_\mathrm{Z} \otimes \uMy = \widetilde{H}^\mathrm{eff}_\mathrm{Y} \otimes \uMz,
 \label{eq:BrownX}
\end{equation}
while the second independent one can be obtained by replacing the subscript $\mathrm{Y}$ by $\mathrm{X}$ everywhere. After substituting the expression for the effective field and the Taylor expansion of the magnetization components in powers of $\uImr$, Brown's equations become Taylor series themselves. By collecting the terms of the same order in $\uImr$, we get a chain of coupled equations for Taylor-expansion coefficients $m_\mathrm{X/Y}^{(1)}$, $m_\mathrm{X/Y}^{(2)}$, etc. For example, the equations in the first order read
\begin{eqnarray}
 (h + \uLeO^2 q^2 + \uyq^2 ) \umIy  + \uxq\uyq\umIx & = & \uAy - \uyq \uzq \uIm, \nonumber \\
 (h + \uLeO^2 q^2 + \uxq^2 ) \umIx  + \uxq\uyq\umIy & = & \uAx - \uxq \uzq \uIm,  \nonumber
\end{eqnarray}
where $h=H/(\gamma_B M_0)$ is the dimensionless field, 
$\uAx=\udx \otimes \udz \otimes \uIk$,
$\uAy=\udy \otimes \udz \otimes \uIk$,
and the dimensionless components of the $\vec{q}$ direction vector are $\{ \uxq, \uyq, \uzq\}=\vec{q}/q$. Quantities $\uLeO$ and $q$ are still carrying dimensions, but their product is dimensionless. These linear equations are solved by 
\begin{eqnarray}
 \umIx & = & \frac{\uAx(h_q + \uyq^2)- \uxq (\uAy \uyq + h_q\uzq \uIm)}
                   {h_q(h_q + \uxq^2 + \uyq^2)}, \label{eq:mx1} \\
 \umIy & = & \frac{\uAy(h_q + \uxq^2)- \uyq (\uAx \uxq + h_q\uzq \uIm)}
                   {h_q(h_q + \uxq^2 + \uyq^2)}, \label{eq:my1} 
\end{eqnarray}
where 
$h_q= h + \uLeO^2 q^2$. When both the exchange interaction and the anisotropy are neglected ($\uAx=0$, $\uAy=0$, $h_q=h$), these expressions coincide with the first-order solution by Schl\"omann\cite{S71}. Otherwise they coincide with the first order solution\cite{michels2013}, which is extensively used at present as a basis for magnetic SANS, except that now we have an explicit expression for $\uAx$ and $\uAy$ via the magnitude and the direction fields of the local uniaxial anisotropy.

In the second order (as well as higher orders), the equations are also linear and differ from the first-order ones only by their right hand side, which now contains sums over the lower-order solutions:
\begin{eqnarray}
 (h + \uLeO^2 q^2 + \uyq^2 ) \umIIy  + \uxq\uyq\umIIx & = & F_\mathrm{Y} + \uyq \uzq F_\mathrm{Z}, \nonumber \\
 (h + \uLeO^2 q^2 + \uxq^2 ) \umIIx  + \uxq\uyq\umIIy & = & F_\mathrm{X} + \uxq \uzq F_\mathrm{Z}.  \nonumber
\end{eqnarray}
Their solutions are also similar,
\begin{eqnarray}
 \umIIx & = & \frac{F_\mathrm{X}(h_q + \uyq^2)- \uxq (F_\mathrm{Y} \uyq - h_q\uzq F_\mathrm{Z})}
                   {h_q(h_q + \uxq^2 + \uyq^2)}, \label{eq:mx2} \\
 \umIIy & = & \frac{F_\mathrm{Y}(h_q + \uxq^2)- \uyq (F_\mathrm{X} \uxq - h_q\uzq F_\mathrm{Z})}
                   {h_q(h_q + \uxq^2 + \uyq^2)}. \label{eq:my2} 
\end{eqnarray}

The special functions are
\begin{eqnarray}
 F_\mathrm{X} & =  & \udx\otimes\uIk\otimes(2 \udz \otimes \uIm + \udx \otimes \umIx + \udy \otimes \umIy) - \nonumber \\
 & & \underline{q\uLeO\uxq\uIe}\otimes\underline{q\uLeO\uxq\umIx} - 
     \underline{q\uLeO\uyq\uIe}\otimes\underline{q\uLeO\uyq\umIx} - \nonumber \\ 
 & &    \underline{q\uLeO\uzq\uIe}\otimes\underline{q\uLeO\uzq\umIx} - \udz\otimes\udz\otimes\uIk\otimes\umIx -
 \nonumber \\
 & &  \uIe \otimes \underline{\uLeO^2 q^2 \umIx}- \nonumber \\
 & & \uIm\otimes\underline{\uLeO^2 q^2 \umIx+\uxq(\uzq\uIm +\uxq \umIx + \uyq \umIy)} + \nonumber \\
 & & \umIx\otimes\underline{\uLeO^2 q^2 \uIm+\uzq(\uzq\uIm +\uxq \umIx + \uyq \umIy)}, \nonumber
\end{eqnarray}
and a similar expression is obtained for $F_\mathrm{Y}$ with the $X$ and $Y$ subscripts as well as the 
functions $\uxq$ and $\uyq$ interchanged. The function $F_\mathrm{Z}= -\umIIz$ is defined by Eq.~(\ref{eq:Fz}).

Just to give a simpler example: if the effects of inhomogeneous anisotropy and exchange are neglected (by putting $\uIk=0$ and $\uIe=0$) and the expressions for $\umIx$ and $\umIy$ are substituted, the special functions are 
\begin{equation}
 F_\mathrm{X/Y/Z} (\vec{q}) =
 \sum\nolimits_{\vec{q}^\prime}
 \frac{\widetilde{I}(\vec{q}^\prime)\widetilde{I}(\vec{q}-\vec{q}^\prime)}{u_\vec{q} u_{\vec{q}-\vec{q}^\prime}}
 f_\mathrm{X/Y/Z}(\vec{q}^\prime,\vec{q}-\vec{q}^\prime) \nonumber
\end{equation}
with
\begin{eqnarray}
 f_\mathrm{X} & = & - h \uxqp \uzqp \uuqmqp - \nonumber \\
 & & (\uzqp^2 (h + \uLeO^2 q^{\prime2}) + \uuqp  \uLeO^2 q^{\prime2}) \uxqmqp \uzqmqp , \nonumber \\
 f_\mathrm{Y} & = & - h \uyqp \uzqp \uuqmqp - \nonumber \\
 & & (\uzqp^2 (h + \uLeO^2 q^{\prime2}) + \uuqp  \uLeO^2 q^{\prime2}) \uyqmqp \uzqmqp , \nonumber \\
 f_\mathrm{Z} & = & \frac{1}{2} \uzqp \uzqmqp (\uxqp \uxqmqp + \uyqp \uyqmqp), \label{eq:fz}
\end{eqnarray}
where $\uuq=h_q + \uxq^2 + \uyq^2$. Expression (\ref{eq:fz}) for $f_\mathrm{Z}$ is valid even if  inhomogeneous exchange and anisotropy are present. If we further neglect the effects of exchange (by putting $\uLeO=0$), the solutions for $\umII$ coincide exactly with those obtained by Schl\"omann\cite{S71}. 

These analytical calculations complete the second-order solution of the micromagnetic problem of a weakly inhomogeneous magnetic material under the influence of an externally applied magnetic field. Let us now proceed with the evaluation of the ensuing magnetic SANS cross sections. 

\section{Model for defects and their averaging}
\label{sec:averag}

The theory of magnetic SANS relates to experiment in a similar way as the theory of the approach to magnetic saturation, being a microscopic theory for a macroscopic measurement. The micromagnetic analysis of the previous section allows us to express the magnetization Fourier image at a specified magnetic field via those of the inhomogeneous saturation magnetization, anisotropy, and exchange. The latter, however, are usually unknown for a specific piece of magnetic material. In fact, it is realistic to assume that the inhomogeneity functions ($\uIm$, $\uIk$, and $\uIe$) are random processes, having specific realizations in each representative volume into which a macroscopic magnet is subdivided. Then, the magnetic SANS cross section, resulting from the scattering of the neutron beam off the macroscopic magnet comprising many representative volumes, can be expressed as an average (both over the random process realization and over the orientation, since the defect realizations in representative volumes are 
also 
randomly oriented).

Also, the inhomogeneities of different material parameters are usually not independent. The underlying physical reasons behind their formation (such as nanocrystallization) imply that the material consists of two or more phases, each having a specific set of magnetic parameters, separated by transition regions (such as grain boundaries). That is why later on we will assume that the inhomogeneity functions are proportional to a universal inhomogeneity function $\widetilde{I}$, describing the material microstructure: $\uIm=\uI$, $\uIk=\kappa \uI$, $\uIe=\epsilon \uI$, where $\kappa,\epsilon \lesssim 1$.

For performing the averaging procedure, it is easiest to start with a specific model for the inhomogeneities. Here, we will consider inhomogeneities which are randomly placed,
\begin{equation}
 I(\vec{r}) = \sum_n f_n(\vec{r}-\vec{p}_n),
\end{equation}
where $\vec{p}_n$ are uniformly-distributed random vectors and the summation is carried out over all the inhomogeneities in the representative volume. The Fourier transform of this function is
\begin{equation}
 \widetilde{I}(\vec{q}) = \sum_n e^{-\imath \vec{q} \vec{p}_n} \widetilde{f}_n(\vec{q}), \label{eq:uI}
\end{equation}
where $\widetilde{f}_n(\vec{q})$ is the Fourier transform of $f_n(\vec{r})$.

We further assume that the inhomogeneities have a Gaussian profile,
\begin{equation}
 f_n(\vec{r}) = a_n e^{-\frac{1}{2}\,\vecT{r} \mat{A} \vec{r}}, \label{eq:Gaussianfr}
\end{equation}
where $a_n$ denotes their (random) amplitude, the subscript $\intercal$ indicates transposition, and the bold capital symbol denotes a square matrix. Moreover, the matrix $\mat{A}$ is assumed to be positive definite and its elements have units of inverse squared length (so that the argument of the exponential is dimensionless). Assuming that the inhomogeneities are much smaller than the representative volume, we can extend the integration limits in the Fourier transform up to infinity to get a simple representation for $\widetilde{f}_n(\vec{q})$,
\begin{equation}
 \widetilde{f}_n(\vec{q}) = \frac{a_n v }{V}
 e^{-\frac{1}{2}\,\vecT{q} \matI{A} \vec{q}}, \label{eq:Gaussianfq}
\end{equation}
where $v=(2\pi)^{3/2}/\sqrt{D}$ is the volume of a single inhomogeneity, $D = \udet \mat{A}$ is the determinant of $\mat A$, and $\matI{A}$ is its inverse.

Since we are going to perform the directional averaging over all the possible inhomogeneity orientations, it is sufficient, without loss of generality, to specify the positive definite matrix $\mat{A}$ in diagonal form. Specifically, to 
consider spheroidal inhomogeneities, we can write the matrix $\mat{A}$ as
\begin{equation}
 \mat{A}= \left(
 \begin{array}{ccc}
  \tau/s^2 & 0 & 0 \\
  0 & \tau/s^2 & 0 \\
  0 & 0 & 1/(\tau s)^2
 \end{array}
  \right), \label{eq:A}
\end{equation}
where $s$ is a real number with units of length specifying the defect size, and $\tau$ is a dimensionless quantity, specifying their shape. The case $\tau=1$ corresponds to spherical inhomogeneities, $\tau\ll1$ to planar and $\tau\rightarrow\infty$ to needle-like elongated defects. The above parametrization is chosen in such a way that the volume $v=(2\pi)^{3/2}s^3$ of a single inhomogeneity is independent of $\tau$.

Now we can explicitly include a rotation matrix $\mat{O}$ into the description of the inhomogeneities. For example, 
we can use a matrix which is parametrized via the spherical angles $\varphi_\mathrm{R}\in[0,2\pi]$ and $\theta_\mathrm{R}\in[0,\pi]$:
\begin{equation}
 \mat{O}= \left(
 \begin{array}{ccc}
  c_\varphi^2 c_\theta + s_\varphi^2 & c_\varphi s_\varphi(c_\theta -1) & c_\varphi s_\theta \\
  c_\varphi s_\varphi(c_\theta -1) & c_\varphi^2  + s_\varphi^2 c_\theta & s_\varphi s_\theta \\
  -c_\varphi s_\theta & -s_\varphi s_\theta & c_\theta
 \end{array}
  \right) ,
\end{equation}
where $c_\varphi=\cos\varphi_\mathrm{R}$, $s_\varphi=\sin\varphi_\mathrm{R}$, $c_\theta=\cos\theta_\mathrm{R}$, and $s_\theta=\sin\theta_\mathrm{R}$; it rotates the direction vector $\{0,0,1\}$ towards the unit-vector 
$\vec{v}=\{c_\varphi s_\theta, s_\varphi s_\theta, s_\theta\}$ and, consequently, has the property 
$\matI{O} \vec{v} = \{0,0,1\}$.

 The quadratic-form matrix $\mat{A}$ in the rotated coordinate system can be represented as $\mat{O}\mat{A}\matI{O} = \mat{O}\mat{A}\matT{O}$, since for the rotation matrix $\matI{O} = \matT{O}$. Similarly, $(\mat{O}\mat{A}\matT{O})^{-1}=\mat{O}\matI{A}\matT{O}$ and $\vecT{q}\vec{p}_n$ in the rotated coordinate system should be replaced by $\vecT{q} \mat{O} \vec{p}_n$. Thus, for the Fourier image of the rotated inhomogeneity function we have
\begin{eqnarray}
 \label{eq:inonavg}
 \widetilde{I}(\vec{q})  & = & \frac{v}{V} \,
 \widetilde{J}(\vec{q}) \,
 e^{-\frac{1}{2}\vecT{q} \mat{O} \matI{A} \matT{O} \vec{q} } , \\
 \widetilde{J}(\vec{q}) & = &  \sum_n a_n e^{-\imath \vecT{q} \mat{O} \vec{p_n}} .
\end{eqnarray}
Besides the averaging over the full range of the rotation angles $\varphi_\mathrm{R}$ and $\theta_\mathrm{R}$, the expressions containing $\widetilde{I}(\vec{q})$ will need to be averaged over the random defect positions $\vec{p}_n$. The function $\uI$ depends on these positions only via the factor $J$. In order to learn how to compute the configurational average of this function, consider its mean-squared value:
\begin{equation}
 \left\langle
 |
    \widetilde{J}(\vec{q})
 |^2
 \right\rangle =
 \left\langle
    \sum_n \sum_{n^\prime} a_n a_{n^\prime} e^{\imath \vecT{q^\prime} \mat{O} (\vec{p}_n-\vec{p}_{n^\prime})}
 \right\rangle = N \langle a_n^2 \rangle , \nonumber
\end{equation}
where $N$ is the number of defects in the representative volume. This is because the averaging of the exponent in the last expression yields a Kronecker delta. The summation can then be easily performed. Similarly, it is possible to show that the various $m$-products of $\widetilde{J}$, such as the triple product $|\widetilde{J}(\vec{q})|^2 \, \mathrm{Re} J(\vec{q})$ or the quadruple product $|\widetilde{J}(\vec{q})|^4$, average over various defect configurations to $N \langle a_n^m \rangle$, which is independent of $\vec{q}$.

Finally, we will assume that the direction of the local anisotropy axis is independent of the particle shape. This means that all the expressions for the SANS cross sections will need to be averaged over the random anisotropy direction as well.

\section{Second-order magnetic SANS cross sections}
\label{sec:2ndorder}

As we have seen in Sec.~\ref{sec:sanscross}, the magnetic SANS cross sections depend on the squared magnetization Fourier components. Since the magnetization components start with the first order in $\uI$, the lowest order terms in the cross sections will be of second order. Let us compute these terms.

\begin{figure}
\includegraphics[scale=0.40]{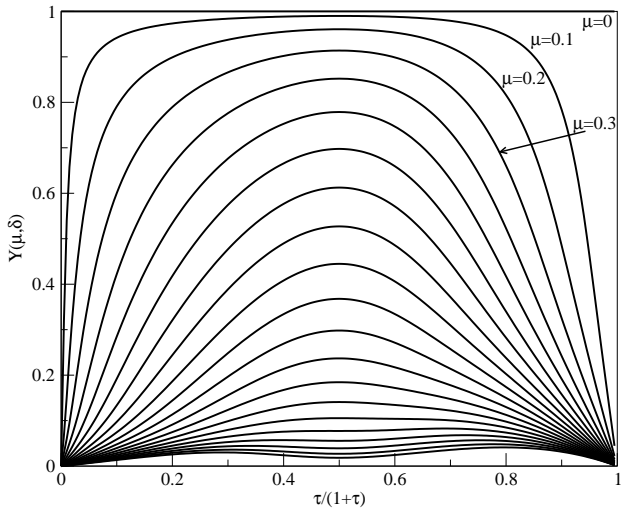}
\caption{\label{fig:upsilon}Dependence of the mean-squared inhomogeneity function 
$\langle \uI^2 \rangle = N\langle a_n^2\rangle (v/V)^2 \Upsilon (q s,\tau)$ on the inclusion shape
$\tau$ for different values of $\mu=q s$ in the range from 0 to 2 in equal steps of $0.1$. $\langle \uI^2 \rangle$ has an extremum at $\tau = 1$, corresponding to spherical defects. The left side of the plot corresponds to planar defects, 
while the right side to needle-like ones.}
\end{figure}

For simplicity, we assume that the magnitude of the anisotropy inhomogeneities is related to the magnitude of the saturation-magnetization inhomogeneities by a factor $\uIk=\kappa\uI$, and also that the anisotropy direction is constant inside each inclusion (but randomly oriented in different ones), so that $\udx=\delta \cos\varphi_\mathrm{A}\sin\theta_\mathrm{A}$, $\udy=\delta \sin\varphi_\mathrm{A}\sin\theta_\mathrm{A}$, and $\udz=\delta \cos\theta_\mathrm{A}$. Then, substituting the magnetization components, Eqs.~(\ref{eq:mx1}) and (\ref{eq:my1}), into the expressions for the parallel [Eq.~(\ref{eq:sigmamagpar})] and perpendicular [Eq.~(\ref{eq:sigmamagperp})] magnetic SANS cross sections, and averaging over the anisotropy directions,
\begin{equation}
 \langle F \rangle_\mathrm{A}=
     \frac{1}{4\pi}\int_0^{2\pi}\int_0^\pi F \sin{\theta_\mathrm{A}}\ud \varphi_\mathrm{A}\ud\theta_\mathrm{A},
\end{equation}
we get
\begin{eqnarray}
 \frac{\ud \Sigma^\parallel_\mathrm{M}}{\ud \Omega} & = &
8\pi^3 V b_\mathrm{H}^2 M_0^2 \langle \uI^2 \rangle \frac{\kappa^2}{15 h_q^2} , \label{eq:o2par} \\
\frac{\ud \Sigma^\perp_\mathrm{M}}{\ud \Omega} & = &
8\pi^3 V b_\mathrm{H}^2 M_0^2 \langle \uI^2 \rangle
\left[ \frac{\kappa^2 \cos^2\alpha}{15(h_q + \sin^2\alpha)^2} + \frac{\kappa^2 }{15 h_q^2} + \right. \nonumber \\
& & \left. \frac{(3+4h_q -\cos2\alpha)\sin^{2}2\alpha}{8 (h_q + \sin^2\alpha)^2} \right] \label{eq:o2perp},
\end{eqnarray}
where we have introduced an angle in the plane of the detector ($q_\mathrm{X}=0$) for the perpendicular cross section $q_\mathrm{Z}=q\cos\alpha$, $q_\mathrm{Y}=q\sin\alpha$ and angular brackets stand for the directional averaging over the representative volume orientations ($\varphi_\mathrm{R}$, $\theta_\mathrm{R}$). For Gaussian defects, Eqs.~(\ref{eq:Gaussianfr})$-$(\ref{eq:A}), this averaging can be performed analytically, yielding $\langle \uI^2 \rangle = N\langle a_n^2\rangle (v/V)^2 \Upsilon (q s,\tau)$ with
\begin{equation}
 \Upsilon (\mu,\tau) =
 \begin{cases}
  e^{-\frac{\mu^2}{\tau}} \frac{\sqrt{\pi}}{2 \mu} \sqrt{\frac{\tau}{1-\tau^3}} 
  \mathrm{Erfi}(\mu \sqrt{\frac{1-\tau^3}{\tau}}) & \tau<1  \\
  e^{-\mu^2} & \tau=1 \\
  e^{-\frac{\mu^2}{\tau}} \frac{\sqrt{\pi}}{2 \mu} \sqrt{\frac{\tau}{\tau^3-1}} 
  \mathrm{Erf}(\mu \sqrt{\frac{\tau^3-1}{\tau}}) & \tau>1
 \end{cases},
\end{equation}
where $\mathrm{Erf}(z)=(2/\sqrt{\pi})\int_0^z e^{-t^2}\ud t$ denotes the error function, and $\mathrm{Erfi}(z)=\mathrm{Erf}(\imath z)/\imath$ is its imaginary counterpart. Dependence of $\Upsilon(\mu, \tau)$ on particle shape at different values of $\mu = q s$ is plotted in Fig.~\ref{fig:upsilon}.

\begin{figure}
\includegraphics[scale=0.51]{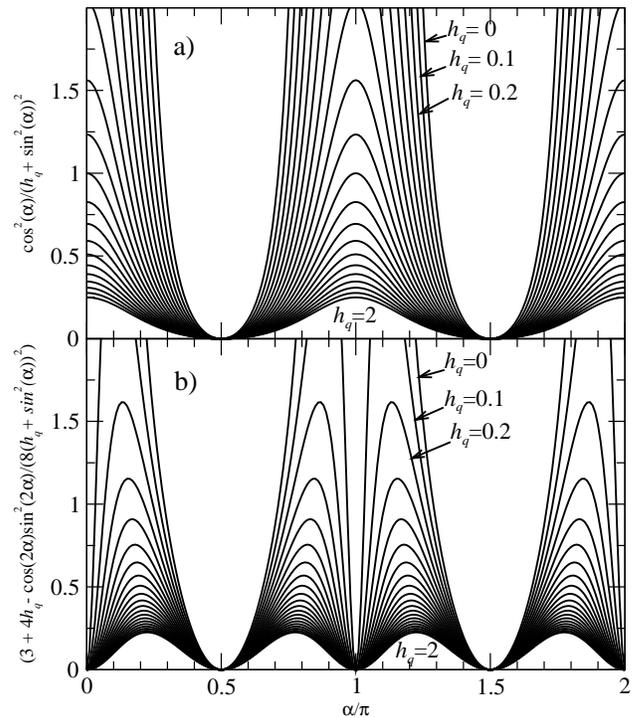}
\caption{\label{fig:perpterms}Angular dependence of the anisotropic terms
in the perpendicular magnetic SANS cross section at different values of $h_q$. (a) displays 
the first and (b) the third term in Eq.~(\ref{eq:o2perp}).}
\end{figure}

The parallel cross section in the second order, Eq.~(\ref{eq:o2par}), is fully isotropic in the detector plane ($q_\mathrm{Z}=0$), while the perpendicular one, Eq.~(\ref{eq:o2perp}), besides the isotropic term $\frac{\kappa^2 }{15 h_q^2}$, contains two terms, which depend on $\alpha$. One of these terms is due to the effect of magnetic anisotropy, while the other is of purely magnetostatic origin. They are plotted in Fig.~\ref{fig:perpterms}.

\begin{figure}
\hspace*{-1cm}\vspace{-0.5cm}\includegraphics[scale=0.8]{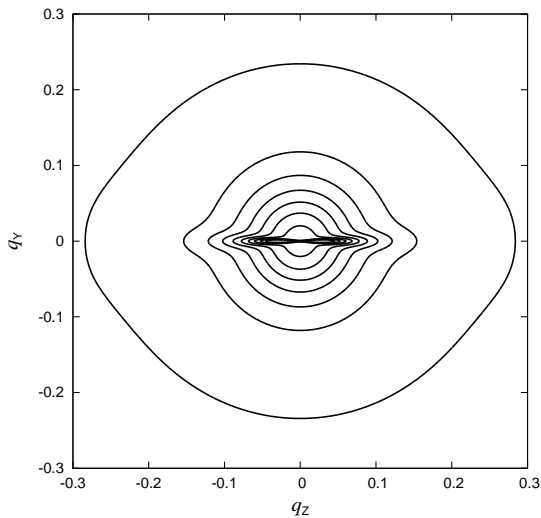}
\caption{\label{fig:ufo}UFO-like magnetic SANS cross section shape at small $h=0.01$ and $q$ in a sample with spherical ($\tau=1$) Gaussian inclusions. The other parameters for this plot are: $\kappa=1$, $s=1$, $\uLeO=1$. The outer contour corresponds to a value of $15$, which increases inwards in steps of $100$. There is a very sharp maximum at the center.}
\end{figure}

Remember that $h_q = h+\uLeO^2 q^2$, which means that $h_q$ takes on values starting with the external field $h>0$ and up to some larger limiting value, dictated by the parameters of the SANS detector. There are two distinct regimes, when $h_q$ is small (small $h$ and small $q$ with respect to the inverse exchange length squared $1/\uLeO$) and when $h_q$ is large (either when $h$ is large, or when $q$ is large for small $h$). In the former regime, the angular dependence of both anisotropic terms displays a similar two-fold angular dependence with sharp maxima along $\alpha=0,\pi$. Together with the isotropic halo, described by the second term in Eq.~(\ref{eq:o2perp}), this gives rise to the recently observed \cite{PGHMM14} UFO-like shape of the SANS image, \cite{wmms99} shown in Fig.~\ref{fig:ufo}.

At large $h_q$, the angular dependence of the first and the third term in Eq.~(\ref{eq:o2perp}) is different. The former is two-fold, while the latter tends asymptotically to $(1-\cos4\alpha)/(4 h_q)$, which has fourfold symmetry. This opens the possibility of separating the anisotropy and the magnetostatic contributions by performing a Fourier analysis of the cross sections at large $h_q$ (either at moderately large $h$ or at the outskirts of the cross section, measured at small $h$).

Regarding the asymptotic $q$-dependence, it is readily verified that both magnetic SANS cross sections vary (for spherical inclusions with $\tau = 1$) as $\sim e^{- s^2 q^2} (s q)^{-4}$, where $s$ denotes the defect size. Other assumptions about the profile of the inclusions, \textit{e.g.}, a sharp interface, may result in different asymptotic dependencies. 

\section{High-order terms in magnetic SANS cross sections}
\label{sec:3rdorder}

The structure of the second-order solutions for the magnetization components, Eqs.~(\ref{eq:mx2}) and (\ref{eq:my2}), is similar to that of the first-order ones, but now magnetostatic effects also contribute to $F_\mathrm{X}$ and $F_\mathrm{Y}$. These functions play the same role in the second-order solutions as the functions $A_\mathrm{X}$ and $A_\mathrm{Y}$ do in the first-order ones, except that they have an additional dependence on the magnitude and the direction of the $\vec{q}$-vector.

The main problem with this (and any other) high-order contribution to physical properties is that it is usually very small and, if lower-order effects are present at the same time, is completely masked by them. On the other hand, analysis of the higher-order effects allows one to extract independently additional information about the system, which the lower-order effects do not provide. Therefore, it is desirable to establish the experimental conditions when the lower-order effects are canceled out and only the higher-order terms contribute, thus, enabling their analysis.

In the present problem this can be achieved by considering the following combination of SANS cross-section values:
\begin{equation}
 \Delta\Sigma^\perp_\mathrm{M} = \left. \frac{\ud \Sigma^\perp_\mathrm{M}}{\ud \Omega}\right|_{\alpha=0} - 
 2 \left. \frac{\ud \Sigma^\perp_\mathrm{M}}{\ud \Omega}\right|_{\alpha= \pi/2} . \label{eq:deltasigma} 
\end{equation}
As can be readily checked from Eq.~(\ref{eq:o2perp}), this combination is exactly zero in second order. This is true both in the presence of anisotropy inhomogeneities $\kappa>0$ and exchange-constant inhomogeneities with an arbitrary (not only Gaussian) spatial profile, since $\langle \uI^2 \rangle$ always depends only on the magnitude of $q$ due to the directional averaging. It is also independent of the assumption that the anisotropy inhomogeneities are related to the saturation-magnetization inhomogeneities by a factor of $\uIk=\kappa\uI$. In other words, the cancellation of the second-order terms in $\Delta\Sigma^\perp_\mathrm{M}$ is a universal property of the SANS cross sections, which is independent of the specific model.

In next significant order (which is the third one), the contributions of $F_\mathrm{X}$ and $F_\mathrm{Y}$ are also canceled and $\Delta\Sigma^\perp_\mathrm{M}$ takes on an especially simple form,
\begin{equation}
 \Delta\Sigma^\perp_\mathrm{M} = 
 32 \pi^3 \, V \, b_\mathrm{H}^2 \, M_0^2 \,  
 \left.\langle F_\mathrm{Z} \uI \rangle \right|_{q_\mathrm{Z}=0},
\end{equation}
where $q=q_\mathrm{Y}$, $F_\mathrm{Z}$ is defined by Eq.~(\ref{eq:Fz}), and the angular brackets denote a triple (configurational, directional, and anisotropy direction) average.

To make the following expressions simpler, let us assume a spherical particle shape ($\tau=1$), which obviates the directional averaging, and, again, assume that $\uIk=\kappa\uI$. Then, averaging is easy to perform,
\begin{equation}
  \Delta\Sigma^\perp_\mathrm{M} = 
 32 \pi^3 \, b_\mathrm{H}^2 \, M_0^2 \, \rho \, v^2 \, \langle a_n^3 \rangle \, [\kappa^2 g_\mathrm{A}(q s)+ g_\mathrm{MS}(q s)],
 \label{eq:sigmaperpthree}
\end{equation}
where $\rho=N/V$ is the defect density. The dimensionless functions $g_\mathrm{A}(\mu)$ and $g_\mathrm{MS}(\mu)$, which also depend on $h$ and $\lambda=\uLeO/s$, are described in the Appendix and plotted in Figs.~\ref{fig:gzkappa} and \ref{fig:gz}. The remaining integrals in these functions are due to the convolution embedded in the definition of the function $F_\mathrm{Z}$.
\begin{figure}
\centering{\includegraphics[scale=0.47]{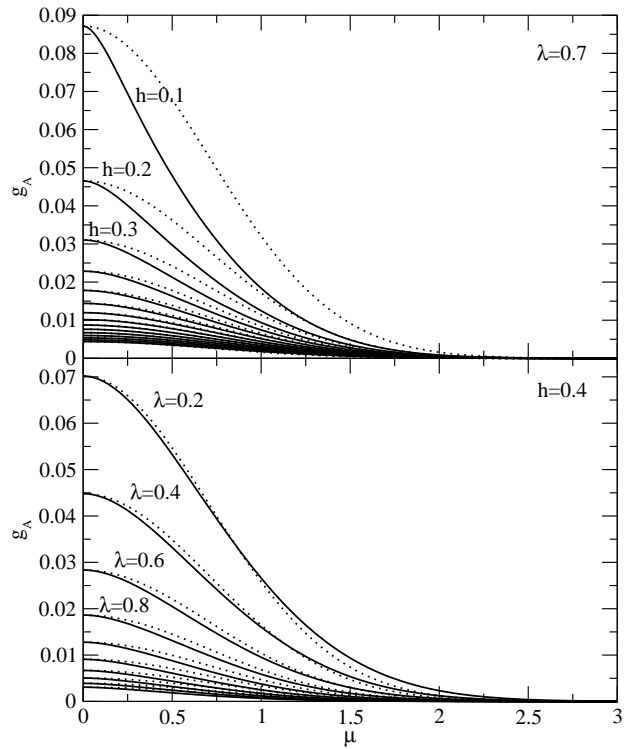}}
\caption{\label{fig:gzkappa}The functions $g_\mathrm{A}(\mu,h,\lambda)$ (solid lines) and their 
approximation by decaying exponentials (dotted lines) for different values of $h$ at fixed $\lambda$
(upper plot) and for different values of $\lambda$ at fixed $h$ (lower plot).}
\end{figure}
\begin{figure}
\centering{\includegraphics[scale=0.47]{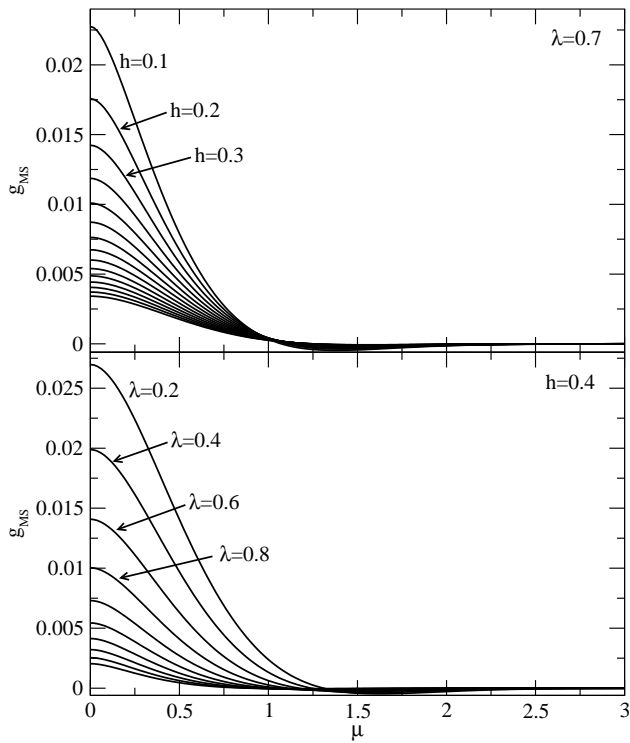}}
\caption{\label{fig:gz}The functions $g_\mathrm{MS}(\mu,h,\lambda)$. The curves correspond to the same values of parameters as those in Fig.~\ref{fig:gzkappa}.}
\end{figure}

The dependence of the third-order perpendicular magnetic difference SANS cross section, Eq.~(\ref{eq:sigmaperpthree}), on $\mu$ for the considered spherical Gaussian defect model is mostly a featureless decaying exponential. Only for small values of the externally applied magnetic field $h$ does this dependence become sharper at small values of $\mu$. In the case of a very small amplitude of the anisotropy inhomogeneities $\kappa$, such that the cross section is dominated by the function $g_\mathrm{MS}$, it is possible to have negative values of $\Delta\Sigma^\perp_\mathrm{M}$ for $\mu \cong 1.5$. This does not, of course, imply that the total cross section is negative.

\section{Summary and conclusions}
\label{sec:sum}

We have presented an analytical solution of the micromagnetic problem of a weakly inhomogeneous magnetic material in an applied magnetic field up to the second order in the amplitude of inhomogeneities. On the basis of this solution, we have computed the second-order magnetic SANS cross sections, which, at sufficiently small values of the applied magnetic field $h$, inevitably display a prominent UFO-like shape. It is shown that under very general assumptions in a magnet with arbitrary small inhomogeneities of exchange, anisotropy, and saturation magnetization, a specific combination of the perpendicular SANS cross-section values, Eq.~(\ref{eq:deltasigma}), is exactly zero in the second order. The next significant third-order contribution to this combination is also computed here and is non-zero. Detection and analysis of its $q$-dependence should provide a deeper insight into the magnet's microstructure. So far there is no experimental confirmation of this newly predicted effect.

\begin{acknowledgments}
Financial support by the National Research Fund of Luxembourg (Project No.~FNR/A09/01) is gratefully acknowledged.
\end{acknowledgments}

\appendix

\section{The functions $g_\mathrm{A}$ and $g_\mathrm{MS}$}

The functions $g_\mathrm{A}$ and $g_\mathrm{MS}$ appear as the result of computing the average $\langle F_\mathrm{Z} \uI \rangle$ over random defect positions, anisotropy direction, and the orientation of the representative volume (if the inclusions are not of spherical shape) with $F_\mathrm{Z}$ defined by Eq.~(\ref{eq:Fz}) and $\uI$ by Eq.~(\ref{eq:uI}). $F_\mathrm{Z}$, however, contains convolutions of the first-order solutions for the magnetization vector field, Eqs.~(\ref{eq:mx1}) and (\ref{eq:my1}), which, in turn, are proportional to $\uI$. For computing these convolutions, it is easiest to approximate the triple summation by a triple integration, according to
\begin{equation}
 \sum\limits_\vec{q} \ldots = \frac{V}{(2\pi)^3} \uiiint \ldots \ud^3\vec{q},
\end{equation}
and integrate over the whole $\vec{q}$ space in a spherical coordinate system. Nevertheless, even in the simplest case of spherical defects (which obviates the directional averages) the full expressions are too complex to be presented here; they are given in the attached Mathematica file\cite{suppmathfile} and plotted in Figs.~\ref{fig:gzkappa} and \ref{fig:gz} (solid lines).

A relatively simple formula can be written for the values of $g_\mathrm{A}$ and $g_\mathrm{MS}$ at $\mu=0$, which reads
\begin{equation}
 \left.g\right|_{\mu=0}=\int\limits_{\sqrt{1+h}}^\infty \!\!\!
 \frac{e^{-(p^2-1-h)/\lambda^2} u(p) \sqrt{p^2-1-h}}{2\sqrt{2\pi}\lambda^3}
 \ud p,
\end{equation}
where the functions $u(p)$ are given by
\begin{eqnarray}
 u_\mathrm{A} & = & \frac{
 \frac{p(3p^2-1)}{(p^2-1)^2}+\coth^{-1}p
 }{15p^2}, \\
 u_\mathrm{MS} & = & -3 p + (3 p^2 - 1) \coth^{-1}p .
\end{eqnarray}
Also, a simple closed-form asymptotic expression for $g_\mathrm{A}$ at large $h$ can be obtained,
\begin{equation}
 g_\mathrm{A}(h\gg1) = \frac{e^{-3\mu^2/4}}{30\sqrt{2} \, h^2}.
\end{equation}

%

\end{document}